\documentstyle[11pt]{article}
\global\arraycolsep=1pt 
\input{tcilatex}
\begin{document}

\font\cmss=cmss10 \font\cmsss=cmss10 at 7pt \hfill \hfill 
CERN-TH/99-234

\hfill hep-th/9908014

\vspace{10pt}

\begin{center}
{\large {\bf \vspace{10pt} TOWARDS THE CLASSIFICATION OF CONFORMAL 
FIELD
THEORIES IN ARBITRARY EVEN DIMENSION}}

\vspace{10pt}

\bigskip \bigskip

{\sl Damiano Anselmi}

{\it CERN, Theory Group, CH-1211, Geneva 23, Switzerland}
\end{center}

\vskip 2truecm

\begin{center}
{\bf Abstract}
\end{center}

\vspace{4pt} 
I identify the class of even-dimensional conformal field
theories that is most similar to two-dimensional conformal field 
theory.
In this class the formula,
elaborated recently, for
the irreversibility of the renormalization-group flow
applies also to massive flows.
This implies a prediction for the ratio between the
coefficient of the Euler density in the trace anomaly
(charge $a$) and the
stress-tensor two-point function (charge $c$). 
More precisely, the trace anomaly in external gravity is quadratic 
in the Ricci tensor and the Ricci scalar and contains a unique
central charge. I check the prediction in detail in four, six and 
eight dimensions, and then in arbitrary even dimension. 
\vskip 1truecm

{Pacs: 11.25.H; 11.10.Gh; 11.15.Bt; 11.40.Ex; 04.62.+v}
\vfill\eject

Four-dimensional conformal field theories have two central charges, 
$c$
and $a$, defined by the trace anomaly in a gravitational background. 
The
charge $c$ multiplies the conformal invariant $W_{\mu \nu \rho \sigma 
}^{2}$
(square of the Weyl tensor) and is the coefficient of the two-point 
function
of the stress tensor. The quantity $a$ multiplies the Euler density 
${\rm G}
_{4}=\varepsilon_{\mu\nu\rho\sigma}\varepsilon^{\alpha\beta\gamma\delta}
R^{\mu\nu}_{\alpha\beta}R^{\rho\sigma}_{\gamma\delta}$. 
A third term, $\Box R$, is multiplied by a coefficient 
$a^{\prime }$: 
\begin{eqnarray}
\Theta =\frac{1}{(4\pi )^{2}}\left[ -cW^{2}+\frac{a}{4}{\rm 
G}_{4}-\frac{2}{3
}a^{\prime }\Box R\right] ,  \label{pic}
\end{eqnarray}
where $c=\frac{1}{120}(N_{s}+6N_{f}+12N_{v})$, $a=\frac{1}{360}
(N_{s}+11N_{f}+62N_{v})$ for free field theories of $N_{s,f,v}$ real
scalars, Dirac fermions and vectors, respectively.

In higher, even dimension $n$ the trace anomaly contains 
more terms, which can however be 
grouped into the same three classes as in four dimensions. 
Several terms are exactly invariant under conformal transformations
and are not total 
derivatives. They generalize $W^{2}$.
The constants in front of these terms will be denoted 
collectively  by $c$.  
One of such central charges, in particular, is related 
to the stress-tensor 
two-point function. It multiplies an invariant 
of the form $W_{\mu \nu \rho \sigma }\Box 
^{n/2-2}W^{\mu
\nu \rho \sigma }+{\cal O}(W^{3})$.
The central charge $a$ is always unique and multiplies the Euler 
density ${\rm G}_{n}$, which is not
conformally invariant, but a non-trivial total derivative.
Finally, the constants in front of the trivial total derivatives, 
which
generalize $\Box R$, will be 
collectively denoted by $a'$.

Only in two dimensions \cite{zamolo} has the trace anomaly a
unique term, the Ricci scalar $R$. In some sense,
we can say that ``$c=a=a^{\prime }$'' there. It 
is
natural to expect that there exists a special
class of higher-dimensional 
conformal
field theories that is most similar to two-dimensional conformal field
theory. This class will have to be identified by a universal 
relationship
between the central charges $c$, $a$ and $a'$.

The main purpose of this paper is to identify this class of
conformal theories, collecting present knowledge and offering further
evidence in favour of the statement. I first
use the sum rule of refs. 
\cite
{athm,at6d} for the irreversibility of the renormalization-group flow 
to
derive a quantitative prediction from this idea, namely the ratio
between the coefficient $a_{n}$ of the Euler density ${\rm G}_{n}$
and the coefficient $c_{n}$ of the invariant $W_{\mu \nu \rho \sigma 
}\Box 
^{n/2-2}W^{\mu
\nu \rho \sigma }+{\cal O}(W^{3})$
(or, which is the same, the constant in front 
of the stress-tensor two-point function). 
Secondly, I argue that the conformal field theories of our special
class are also those whose trace anomaly in external gravity is
quadratic in the Ricci tensor and Ricci scalar. This property relates 
unambigously the central charges $c$ to the unique central charge $a$ 
and, in particular, should agree with
the ratio $c_{n}/a_{n}$ found using the irreversibility of the 
RG flow.

I then proceed to check
the prediction. This is first done in detail in four, six and eight 
dimensions and
then extended to the general case. The results are also a very 
non-trivial 
test of the ideas of refs. \cite{athm,at6d} about the irreversibility 
of the RG flow.

\medskip

I recall that in \cite{n=4,n=2} it was shown that
in four dimensions there is a ``closed limit'',
in which the
stress-tensor operator product expansion (OPE) closes with a finite 
number
of operators up to the regular terms. The idea of this 
limit was suggested by a powerful theorem,
due to Ferrara, Gatto and Grillo \cite{grillo} and to Nachtmann 
\cite{nach}, on the
spectrum of anomalous dimensions of the higher-spin currents 
generated by the OPE, which
follows from very general principles (unitarity) and is therefore 
expected
to hold in arbitrary dimension.

When $c=a$ \cite{n=4}, OPE closure is achieved in a way that is 
reminiscent of
two-dimensional conformal field theory, with the stress tensor and the
central extension. Instead, when $c\neq a$ the algebraic structure is
enlarged and contains spin-1 and spin-0 operators, yet in finite 
number.
Therefore, the subclass of theories we are interested in is 
identified, in four
dimensions, by the equality of $c$ and $a$ and the closed limit.
Secondly, it is well known that $
\Theta $ vanishes on Ricci-flat metrics when $c=a$ in four 
dimensions. A
closer inspection of (\ref{pic}) shows that actually $\Theta $ is 
{\it 
quadratic} in the Ricci tensor and the Ricci scalar.
We are led to conjecture that the subclass of ``$c=a$''-theories 
in arbitrary even dimension are 
those that have a trace anomaly quadratic in the Ricci tensor and the 
Ricci scalar.

Summarizing, in arbitrary even dimension greater than 2 we can 
distinguish the following important
subclasses of conformal field theories:

{\it i}) The ``closed'' theories, when the quantum conformal algebra, 
i.e. the
algebra generated by the singular terms of the stress-tensor OPE, 
closes
with a finite number of operators. They can have $c=a$ \cite{n=4}, 
but also 
$c\neq a$ \cite{n=2}.

{\it ii}) The $c=a$-theories, whose trace anomaly is quadratic in the 
Ricci
tensor and the Ricci scalar. They can be either closed or open.

{\it iii}) The closed $c=a$-theories, which exhibit the highest 
degree of
similarity with two-dimensional conformal field theory.

\medskip

While the equality $c=a$ is a restriction on the set of conformal 
field
theories, the equality of $a$ and $a^{\prime }$ is not. In refs. \cite
{athm,at6d} the equality $a=a^{\prime }$ was studied in arbitrary even
dimension $n$, leading to the sum rule \begin{eqnarray}
a_{n}^{{\rm UV}}-a_{n}^{{\rm 
IR}}={\frac{1}{2^{{\frac{3n}{2}}-1}n\,n!}}\int {\rm 
d}^{n}x\,|x|^{n}\langle \Theta (x)\,\Theta (0)\rangle ,  
\label{sum}
\end{eqnarray}
expressing the total renormalization-group (RG)\ flow of the central 
charge $
a_{n}$, induced by the running of dimensionless couplings. This 
formula was checked to the fourth-loop order included in the most 
general renormalizable theory in four \cite{athm} and six \cite{at6d} 
dimensions.
No restriction on the central charges $c$ and $a$
is required here. The charge $a_{n}$ is normalized so
that the trace anomaly reads \[
\Theta =a_{n}{\rm G}_{n}=a_{n}(-1)^{\frac{n}{2}}\varepsilon _{\mu 
_{1}\nu
_{1}\cdots \mu _{\frac{n}{2}}\nu _{\frac{n}{2}}}\varepsilon ^{\alpha
_{1}\beta _{1}\cdots \alpha _{\frac{n}{2}}\beta 
_{\frac{n}{2}}}\prod_{i=1}^{
\frac{n}{2}}R_{\alpha _{i}\beta _{i}}^{\mu _{i}\nu _{i}}
\]
plus conformal invariants and trivial total derivatives.

As it was explained in the introduction of \cite{athm}, the
arguments of \cite{athm,at6d} do
not necessarily apply to flows generated by super-rinormalizable 
couplings and mass terms. (In
general, the effect of masses can be included straightforwardly 
\cite{noi2}.) The sum rule (\ref{sum}) measures the effect of the 
dynamical RG
scale $\mu $ in lowering the amount of massless degrees of freedom of 
the
theory along the RG\ flow.

The basic reason why massive flows behave differently is that in a 
finite
theory Duff's identification \cite{duff} $a^{\prime }=c$ is 
consistent (but
not unique), while along a RG flow the only consistent identification 
is $
a^{\prime }=a$, as shown in \cite{athm}. Divergences are crucial in
discriminating between the two cases. A flow induced by divergences 
cannot, in general, be assimilated to a flow induced by explicit 
(``classical'') scales.

Repeating the arguments of \cite{athm,at6d} in two dimensions, we 
would
come to the
same conclusion as in higher dimensions: that the sum rule 
(\ref{sum}) works for RG flows and
not necessarily for massive ones. The point is, nevertheless, that the
two-dimensional version of (\ref{sum}), due to Cardy \cite{cardysum},
is universal; in particular, it does work
for massive flows. It is therefore compulsory to understand in what
cases the domain of validity of our sum rule (\ref{sum}) is similarly
enhanced in higher dimensions. This property identifies the special 
class of theories we are looking for.

The arguments and explicit checks that we now present show that this
enhancement takes place in the subclass of theories with $c=a$ 
(classes {\it ii}
and {\it iii} above), because of the higher similarity with the 
two-dimensional theories.

The two relevant terms of the trace anomaly are \[
\Theta =a_{n}{\rm G}_{n}-{\frac{c_{n}(n-2)\,\left( 
{\frac{n}{2}}\right) !}{
4(4\pi )^{\frac{n}{2}}(n-3)\,(n+1)!}}W\Box ^{{\frac{n}{2}}-2}W+\cdots 
, \]
where \[
c_{n}=N_{s}+2^{{\frac{n}{2}}-1}(n-1)N_{f}+{\frac{n!}{2\left[ \left( 
{\frac{n
}{2}}-1\right) !\right] ^{2}}}N_{v} \]
is the value of the central charge $c$ for free fields, and in 
arbitrary
dimension $n$. $N_{v}$ denotes the number of $\left( n/2-1\right) 
$-forms.
This calculation is done in ref. \cite{iera}, section 9, starting 
from the
stress-tensor two-point function.

Massive flows have been considered, among other things,
by Cappelli et al. in \cite{cap}. An explicit computation for free 
massive scalar fields
and fermions gives \cite{cap} \begin{eqnarray}
\int {\rm d}^{n}x\,|x|^{n}\,\langle \Theta (x)\,\Theta (0)\rangle 
={\frac{
c_{n}\,\left( {\frac{n}{2}}\right) !}{\pi ^{\frac{n}{2}}\,(n+1)}}.
\label{sum2}
\end{eqnarray}
Repeating the computation for massive vectors, or $\left( 
n/2-1\right) $
-forms, is problematic in the UV. However, the relative coefficient 
between
the scalar and fermion contributions is sufficient to show that the 
result
is proportional to $c_{n}$ and not $a_{n}$.

Our prediction is that in the special $c=a$-theories the sum rule 
(\ref
{sum}) should reproduce (\ref{sum2}) for massive flows, which means 
\begin{eqnarray}
c_{n}=a_{n}{\frac{2^{{\frac{n}{2}}-1}(4\pi 
)^{\frac{n}{2}}\,n\,(n+1)!}{
\left( {\frac{n}{2}}\right) !}.}  \label{c=a}
\end{eqnarray}
The trace anomaly therefore has the form \begin{eqnarray}
\Theta ={a}_{n}\left( {\rm 
G}_{n}-{\frac{2^{{\frac{n}{2}}-3}n(n-2)}{n-3}}
W\Box ^{n/2-2}W\right) +\cdots   \label{prediction}
\end{eqnarray}

Formula (\ref{c=a}) is the generalized version of the relation $c=a$. 
It
is uniquely implied by the requirement that $\Theta $ be quadratic in 
the
Ricci tensor and Ricci curvature. This condition fixes all the central
charges of type $c$ in terms of $a_{n}$, not only the constant 
$c_{n}$ in
front of the stress-tensor two-point function. These further 
relationships
are not important for our purposes.

In four dimensions the combination between the parenthesis in 
(\ref{prediction})
is indeed quadratic in the Ricci tensor:
\[
\frac{{\rm G}_{4}}{4}-W^{2}=-2R_{\mu \nu }^{2}+\frac{2}{3}R^{2}.
\]
I stress that this is a non-trivial check of the prediction that 
formula (\ref{sum}) correctly describes massive flows when $c=a$.

In higher dimensions the check is less straightforward, owing to the 
high
number of invariants. Using the results of Bonora et al. from 
\cite{bonora} (see also \cite{deser}),
where the terms occurring in the trace anomaly were classified in six
dimensions, we can perform a second non-trivial
check of our prediction. The conformal invariants are three:
\begin{eqnarray*}
I_{1} &=&W_{\mu \nu \rho \sigma }W^{\mu \alpha \beta \sigma 
}W_{~\alpha
\beta }^{\nu \quad \rho},\quad I_{2}=W_{\mu \nu \rho \sigma }W^{\mu 
\nu
\alpha \beta }W_{\alpha \beta }^{\quad ~\rho \sigma }, \\
I_{3} &=&W_{\mu \alpha \beta \gamma }\left( \Box ~\delta _{\nu }^{\mu
}+4R_{\nu }^{\mu }-\frac{6}{5}R~\delta _{\nu }^{\mu }\right) W^{\nu 
\alpha
\beta \gamma },
\end{eqnarray*}
and the general form of the trace anomaly is \[
\Theta =a_{6}{\rm G}_{6}+\sum_{i=1}^{3}c^{(i)}I_{i}+{\rm t.t.d.},
\]
where ``t.t.d.'' means ``trivial total derivatives'' (as opposed to 
${\rm G}
_{6}$, which is a non-trivial total derivative). Our notation differs 
from
the one of \cite{bonora} in the signs of $R_{\mu \nu }$ and $R$. More
importantly, the invariant $I_{3}$ differs from the invariant $M_{3}$ 
of \cite{bonora} and other references \cite{armeni1}, the latter
containing a spurious contribution proportional to ${\rm G}_{6}$ (see 
also \cite{at6d}, section 3), as well as a linear combination of 
$I_{1}$ 
and $
I_{2}$. Precisely, we find \[
M_{3}=\frac{5}{12}{\rm 
G}_{6}+\frac{80}{3}I_{1}+\frac{40}{3}I_{2}-5I_{3}.
\]
Finally, our $I_{3}$ differs from the expression of ref. 
\cite{armeni},
formula (19), by the addition of t.t.d.'s, which, however, can be
consistently omitted for our purposes.

In \cite{bonora} it is pointed out that there exists a simple 
combination of
the four invariants ${\rm G}_{6}$ and $I_{1,2,3}$, which reads 
\begin{eqnarray}
{\cal J}_{6} &=&R_{\mu \nu }\Box R^{\mu \nu }-\frac{3}{10}R\Box 
R-RR_{\mu
\nu }R^{\mu \nu }  \nonumber \\
&&-2R_{\mu \nu }R_{\rho \sigma }R^{\mu \rho \sigma \nu 
}+\frac{3}{25}R^{3} \nonumber \\
&=&-\frac{1}{24}{\rm G}_{6}-4I_{1}-I_{2}+\frac{1}{3}I_{3}+{\rm t.t.d.}
\label{simple}
\end{eqnarray}
The BPB (Bonora--Pasti--Bregola) term ${\cal J}_{6}$ is precisely the
combination we are looking for. A closer inspection of this 
expression shows
that it is uniquely fixed by the requirement that it be quadratic in 
the
Ricci tensor and Ricci curvature. On the other hand, the requirement 
that $
{\cal J}_{6}$ just vanishes on Ricci-flat metrics is not sufficient 
to fix
it uniquely, in particular it does not imply the relation ``$c=a$'' 
that we
need.

In conclusion, the $c=a$-theories have a unique central charge,
multiplying the BPB invariant ${\cal J}_{6}$, \[
\Theta =-{24~}a_{6}~{\cal J}
_{6},~~~c^{(1)}=96a_{6},~~c^{(2)}=24a_{6},~~c^{(3)}=-8a_{6}, \]
so that $\Theta $ is of the predicted form (\ref{prediction}): 
\begin{eqnarray}
\Theta =a_{6}({\rm G}_{6}-8W\Box W)+\cdots .  \label{che}
\end{eqnarray}

\medskip

Our prediction is meaningful in arbitrary even dimension and
can be checked using the recent work of Henningson and Skenderis 
\cite{skende}, which contains, as I now discuss, an algorithm
to generate precisely 
the invariants ${\cal J}_{n}$'s that we need. It is easy to verify
this 
in four and six dimensions. In six dimensions the result can be read 
from formula (30) of \cite{skende}, taking into account that in 
\cite{skende} the BPB invariant $M_{3}$ is used. A more convenient 
decomposition of the anomaly into Euler density and
conformal invariants is the last equality of (\ref{simple}), leading
directly to (\ref{che}).
It is therefore natural to expect that the algorithm
of \cite{skende} answers our question and constructs the
invariants ${\cal J}_{n}$'s. I now check agreement with
formula (\ref{prediction}) in arbitrary even dimension.

I begin with $n=8$. The relevant terms of ${\cal J}_{8}$ are \[
{\cal J}_{8}=R_{\mu \nu }\Box ^{2}R^{\mu \nu }-\frac{2}{7}R\Box 
^{2}R+{\cal O
}(R^{3})=\alpha _{8}~{\rm G}_{8}+{\rm c.i.}+{\rm t.t.d.},
\]
$\alpha _{8}$ being the unknown coefficient and ``${\rm c.i.}$'' 
denoting
conformal invariants. On a sphere, in particular, all terms but 
$\alpha _{8}
{\rm G}_{8}$ vanish, so that $\alpha _{8}${\rm \ }can be found by 
evaluating
the integral of ${\cal J}_{8}$: \[
\int_{S^{8}}\sqrt{g}{\cal J}_{8}~{\rm d}^{8}x=768~\alpha _{8}~(4\pi 
)^{4}.
\]
Using \[
W\Box ^{2}W=\frac{10}{3}\left( R_{\mu \nu }\Box ^{2}R^{\mu \nu 
}-\frac{2}{7}
R\Box ^{2}R\right) +{\cal O}(R^{3})+{\rm t.t.d.},
\]
our prediction (\ref{prediction}) is $\alpha _{8}=-1/64.$ Indeed, 
applying
the method of \cite{skende} on a conformally-flat metric with $R_{\mu 
\nu }=\Lambda g_{\mu
\nu },$ we get, after a non-trivial amount of work, \[
{\cal J}_{8}=\alpha _{8}~{\rm G}_{8}=-\frac{1440}{343}\Lambda ^{4},
\]
which gives the desired value of $\alpha _{8}$.

The check can be generalized for arbitrary $n$. The invariant ${\cal 
J}_{n}$
is, up to an overall factor $\beta _{n}$, the coefficient of $\rho 
^{n/2}$
in the expansion of $\sqrt{\det G},$ where \[
G_{\mu \nu }=g_{\mu \nu }+\sum_{k=1}^{n/2}\rho ^{k}g_{\mu \nu 
}^{(k)}+{\cal O
}(\rho ^{n/2}\ln \rho ,\rho ^{n/2+1},\cdots )
\]
and the $\rho $-dependence is fixed by the equations \cite{skende} 
\begin{eqnarray}
{\rm tr}[G^{-1}G^{\prime \prime }]-\frac{1}{2}{\rm tr}[G^{-1}G^{\prime
}G^{-1}G^{\prime }] &=&0,  \label{ske} \\
2\rho (G^{\prime \prime }-G^{\prime }G^{-1}G^{\prime }) &=&(G-\rho 
G^{\prime
}){\rm tr}[G^{-1}G^{\prime }]  \nonumber \\
&&+{\rm Ric}(G)+(n-2)G^{\prime }.  \nonumber
\end{eqnarray}
Precisely, \begin{eqnarray*}
{\frac{1}{\left( {\frac{n}{2}}\right) !}}\left. {\frac{{\rm 
d}^{\frac{n}{2}}
}{{\rm d}\rho ^{\frac{n}{2}}}}{\frac{\sqrt{\det G}}{\sqrt{\det 
g}}}\right|
_{\rho =0} &=&\beta _{n}{\cal J}_{n} \\
&&\!\!\!\!\!\!\!\!\!\!\!\!{\beta _{n}(R_{\mu \nu }\Box 
^{{\frac{n}{2}}-2}R^{\mu \nu }+\alpha _{n}
{\rm G}_{n}+{\rm rest}).}
\end{eqnarray*}

First, we consider metrics with $R_{\mu \nu }=\Lambda g_{\mu \nu }$. 
The
form of the solution and the first equation of (\ref{ske}) read \[
G_{\mu \nu }=u(\rho \Lambda )g_{\mu \nu },\qquad \frac{u^{\prime 
\prime }}{u}
=\frac{1}{2}\left( \frac{u^{\prime }}{u}\right) ^{2}. \]
The second equation of (\ref{ske}) is used to fix the integration 
constants,
with the result \[
u(\rho \Lambda )=\left( 1-\frac{\rho \Lambda }{4(n-1)}\right) ^{2},~~
\beta _{n}{\cal J}_{n}\rightarrow \frac{(-1)^{\frac{n}{2}}~n!~\Lambda 
^{
\frac{n}{2}}}{2^{n}(n-1)^{\frac{n}{2}}\left[ \left( \frac{n}{2}\right)
!\right] ^{2}}. \]
Then, we fix the normalization $\beta _{n}$ by looking for the term $
R_{\mu \nu }\Box ^{\frac{n}{2}-2}R^{\mu \nu }$ (we can set the Ricci
curvature $R$ to zero for simplicity). We write \[
G_{\mu \nu }=g_{\mu \nu }+\frac{1}{\Box }v(\rho \Box )R_{\mu \nu 
}+R_{\mu
\alpha }\frac{1}{\Box ^{2}}y(\rho \Box )R^\alpha_\nu+O(R^{3}), \]
with $v(0)=y(0)=y^{\prime}(0)=0$. We have \[
\beta _{n}=\frac{1}{2\left( \frac{n}{2}\right) !}\left. \frac{{\rm 
d}^{\frac{
n}{2}}x}{{\rm d}t^{\frac{n}{2}}}\right| _{t=0} \]
where $t=\rho \Box $ and $x=y-v^{2}/2$. Integrating ${\cal J}_{n}$ 
over a
sphere, we can convert our prediction (\ref{prediction}) to a 
prediction for $
\beta _{n}$ or \begin{eqnarray}
\left. \frac{{\rm d}^{\frac{n}{2}}x}{{\rm d}t^{\frac{n}{2}}}\right| 
_{t=0}=-
\frac{1}{2^{n-1}\Gamma \left( \frac{n}{2}\right) }.  \label{dera}
\end{eqnarray}
Equations (\ref{ske}) relate $y$, and therefore $x$, to $v$ and imply 
that $
v $ is a Bessel function of the second type:
\begin{eqnarray}
x^{\prime \prime }=-\frac{(v^{\prime })^{2}}{2},\quad \quad 
2tv^{\prime
\prime }-1+\frac{v}{2}-(n-2)v^{\prime }=0.  \label{recu}
\end{eqnarray}
$\beta _{n}$ is a coefficient in the series expansion of the square 
of a
Bessel function of the second type, and is not usually in the 
mathematical
tables. Solving (\ref{recu}) recursively with the help of a 
calculator, we
have checked agreement between (\ref{dera}) and (\ref{recu}) up to 
dimension $1000$.

Our picture and the quantitative agreement with prediction 
(\ref{prediction}
) explain, among other things, the physical meaning of the 
construction of ref. \cite{skende}. Furthermore, the mathematical 
properties of the
invariant ${\cal J}_{n}$, and therefore the identification of $c$ and 
$a$
(in the subclasses of theories {\it ii} and {\it iii} where it 
applies), are a nice counterpart
of the notion of extended ({\it pondered}) Euler density introduced 
in \cite
{at6d}, which explained the identification $a=a^{\prime }$.
The results presented in this paper are a further check of the
ideas of \cite{athm,at6d} and of the picture offered there. These 
are, we
believe, the first steps towards the classification of all conformal 
field
theories.

The set of higher-dimensional quantum field theories, conformal or 
not, is not rich of physical models. Yet, one can consider 
higher-derivative theories, which, despite the issues about unitarity 
(see for example \cite{antoniadis}), are useful
toy-models for our purposes. Here higher-dimensional higher-derivative
theories are 
meant as a convenient laboratory where the results of the present
paper might be applied.

\vskip .2truecm

I thank A. Cappelli for reviving my interest for massive
flows, the organizers of the 4$^{\it th}$ Bologna Workshop on CFT and 
integrable models, D.Z. Freedman and N. Warner for stimulating
conversations on the four-dimensional problem, M. Porrati for drawing 
my
attention to the six-dimensional results of ref. \cite{skende}, and 
finally
L. Girardello and A. Zaffaroni.

\end{document}